\documentclass[prd,aps,twocolumn,nofootinbib,preprintnumbers,superscriptaddress,preprintnumbers,balancelastpage,longbibliography]{revtex4-2}
\usepackage[english]{babel}

\usepackage{amsmath,mathtools,physics,xfrac}
\usepackage{graphicx}
\usepackage{afterpage}
\usepackage{float}
\usepackage{subfigure}
\usepackage{rotating}
\usepackage{multirow}
\usepackage{tabularx}
\usepackage{booktabs}
\usepackage{fancyhdr}
\usepackage[utf8]{inputenc}
\usepackage{theorem}
\usepackage{moreverb}
\usepackage{euscript}
\usepackage{psfrag}
\usepackage{slashed}
\usepackage{mathtools}
\usepackage{makecell}
\usepackage{adjustbox}
\usepackage{dcolumn}
\usepackage{bm}
\usepackage[dvipsnames]{xcolor}
\usepackage{graphics}
\usepackage{hyperref}
\usepackage{chngcntr}
\usepackage{aas_macros}
\usepackage{color,xcolor}
\usepackage[normalem]{ulem}
\usepackage{amssymb}
\hypersetup{
     colorlinks   = true,
     citecolor    = blue,
     urlcolor     = blue,
     linkcolor    = blue
}

\definecolor{darkblue}{rgb}{0.0,0.0,0.75}
\definecolor{darkred}{rgb}{0.6,0.0,0}
\definecolor{darkgreen}{rgb}{0.0,0.6,0.}

\counterwithout{equation}{section}

\newcommand{\eqnref}[1]{Eq.~\eqref{#1}}
\newcommand{\figref}[1]{Fig.~\ref{#1}}

\usepackage{tikz,xcolor,hyperref}

\definecolor{lime}{HTML}{A6CE39}
\DeclareRobustCommand{\orcidicon}{\hspace{-1mm}
	\begin{tikzpicture}
		\draw[lime, fill=lime] (0,0) 
		circle [radius=0.16] 
		node[white] {{\fontfamily{qag}\selectfont \tiny \,ID}};
		\draw[white, fill=white] (-0.0525,0.095) 
		circle [radius=0.007];
	\end{tikzpicture}
	\hspace{-3mm}
}

\foreach \x in {A, ..., Z}{\expandafter\xdef\csname orcid\x\endcsname{\noexpand\href{https://orcid.org/\csname orcidauthor\x\endcsname}
		{\noexpand\orcidicon}}
}


\begin{document}

\title{Ly\texorpdfstring{$\alpha$}{alpha} forest bounds on sterile neutrino production via neutrino self-interactions}

\author{Priyank Parashari\orcidA{}}
\email{ppriyank@usc.edu}
\affiliation{Department of Physics $\&$ Astronomy, University of Southern California, Los Angeles, CA, 90089, USA}

\author{Vera Gluscevic}
\email{verag@ias.edu}
\affiliation{Department of Physics $\&$ Astronomy, University of Southern California, Los Angeles, CA, 90089, USA}
\affiliation{Institute for Advanced Study, 1 Einstein Drive, Princeton, NJ 08540, USA}
\affiliation{Center for Computational Astrophysics, Flatiron Institute, 162 5th Avenue, New York, NY, 10010, USA}
\author{Yue Zhang}
\affiliation{Department of Physics, Carleton University, Ottawa, ON K1S 5B6, Canada}
\author{Simeon Bird}
\affiliation{Department of Physics and Astronomy, University of California Riverside, Riverside, CA 92521}
\author{Mikhail M. Ivanov}
\affiliation{Center for Theoretical Physics – a Leinweber Institute, Massachusetts Institute of Technology, Cambridge, MA 02139, USA}
\affiliation{The NSF AI Institute for Artificial Intelligence and Fundamental Interactions, Cambridge, MA 02139, USA}
\author{Adam He}
\affiliation{Department of Physics $\&$ Astronomy, University of Southern California, Los Angeles, CA, 90089, USA}

	
\begin{abstract}

Sterile neutrinos in the keV mass range have long been considered a well-motivated dark matter (DM) candidate. In this work, we explore a sterile neutrino production mechanism through active neutrino self-interactions in the early universe, assuming that they constitute the full DM abundance. We implement a self-consistent treatment of the sterile-neutrino free streaming and the active-neutrino self-interactions on structure formation, which yield a unique scale-dependent modification to the linear matter power spectrum. We then set bounds on this scenario using a combination of the cosmic microwave background and Ly$\alpha$ forest constraints. Specifically, we utilize the two recent likelihoods derived from eBOSS data: (i) an effective field theory (EFT) based full-shape likelihood and (ii) a compressed likelihood obtained from the PRIYA-simulation emulator. We produce some of the most stringent observational constraints to date on sterile neutrino DM, comparable to the bounds from the most stringent laboratory constraints.

\begin{description}
\item[Keywords]
Cosmology, Dark matter, Sterile neutrino, Self-interaction, Lyman-$\alpha$ Forest
\end{description}

\end{abstract}

\preprint{MIT-CTP/6004}
	
	\maketitle

\section{Introduction}\label{sec:intro}
The nature of dark matter (DM) remains one of the most profound open questions in modern cosmology and particle physics. Despite astrophysical and cosmological observations providing compelling evidence for the existence of DM, its fundamental properties remain unknown\cite{Bertone:2004pz,Bertone:2016nfn,Cirelli:2024ssz,Planck:2018vyg}. A sterile neutrino with mass in the keV range, having no direct interaction with the Standard Model particles but interacting feebly via mixing with active neutrinos, has been proposed as a well-motivated DM candidate~\cite{Gunn:1978gr,Tremaine:1979we,Abazajian:2017tcc}. Unlike the three thermally-produced active neutrinos of the Standard Model, the production of this hypothetical fourth neutrino state involves a non-thermal freeze-in mechanism via active-sterile neutrino oscillations. The sterile neutrino state $\nu_4$ is characterized by its mass $m_4$ and a small mixing angle $\theta$, and its mass eigenstate is defined as a linear combination of a predominantly sterile component $\nu_s$ and a tiny component of active neutrinos $\nu_a$, 
\begin{equation}
|\nu_4\rangle = \cos\theta|\nu_s\rangle + \sin\theta |\nu_a\rangle\,.
\label{eq:_nu_mixing}
\end{equation}
Dodelson and Widrow~\cite{1994PhRvL..72...17D} (DW) demonstrated that the sterile neutrino, whose properties make it a compelling DM candidate, can be produced with the full relic abundance of DM, provided the appropriate combination of mass and mixing angle. In the DW framework, the production through oscillations continues until the active neutrinos decouple from the rest of the particles, leaving behind a relic abundance of sterile neutrino DM. 

However, the original DW mechanism is now ruled out by observations. Most importantly, the same active-sterile oscillations responsible for sterile neutrino production would also lead to radiative decay of sterile neutrino into a neutrino and a photon, and an X-ray line-emission in overdense regions of the universe, such as galaxies and galaxy clusters~\cite{Pal:1981rm}. Even a small mixing angle required to produce the observed DM abundance via DW mechanism in the early universe would generate a detectable X-ray signal at most particle masses; the full range of mixing angles required to produce the relic abundance of DM with masses $m_4 \gtrsim 2$ keV via the original DW mechanism is excluded by X-ray bounds~\cite{Boyarsky:2005us,2012JCAP...03..018W,Horiuchi:2013noa,Perez:2016tcq,2020Sci...367.1465D,Krivonos:2024yvm}. Lower masses are also excluded by various small-scale-structure constraints, as discussed in more detail below \cite{Nadler:2019zrb,Tremaine:1979we,Boyarsky:2008ju,Watson:2011dw,Viel:2013fqw,Baur:2015jsy,Horiuchi:2015qri,Yeche:2017upn,Irsic:2017ixq,DES:2020fxi,Nadler:2021dft,Villasenor:2022aiy,Irsic:2023equ,An:2023mkf}.

New mechanisms for sterile neutrino production beyond the original DW scenario have been proposed in the literature~\cite{Shi:1998km,Shaposhnikov:2006xi, Asaka:2006ek, Bezrukov:2009th, Nemevsek:2012cd, Roland:2014vba, Dror:2020jzy, Nemevsek:2022anh, 2020PhRvL.124h1802D, 2020PhRvD.101k5031K, Kelly:2020aks,Chichiri:2021wvw,Benso:2021hhh,Vogel:2025gan,Dev:2025sah,Akita:2025txo}. One of these mechanisms involves the introduction of self-interactions among active neutrinos, which can boost sterile production at a fixed mixing angle, and relax observational constraints by opening up new parameter space for this DM model~\cite{2020PhRvL.124h1802D, 2020PhRvD.101k5031K,Kelly:2020aks,Chichiri:2021wvw,Benso:2021hhh,Vogel:2025gan}. Neutrino self-interactions are well-motivated in their own as they can naturally arise in extensions of the Standard Model, particularly in context of neutrino mass generation mechanisms, such as models involving Majoron-like states or additional scalar and gauge degrees of freedom~\cite{Chikashige:1980ui,Gelmini:1980re}. More recently, they have also been explored in cosmological studies, as they can affect the Hubble tension~\cite{Kreisch:2019yzn,Blinov:2019gcj,Mazumdar:2020ibx,Lyu:2020lps,Brinckmann:2020bcn,RoyChoudhury:2020dmd,Das:2021guu}. At the same time, neutrino self-interaction scenarios have also been extensively explored using both laboratory experiments~\cite{Barger:1981vd,ALEPH:2005ab,2018PhRvD..97g5030B,Agostini:2015nwa,Blum:2018ljv,Berryman:2018ogk,Chauhan:2018dkd,Blinov:2019gcj,Brdar:2020nbj,Lyu:2020lps,PIENU:2020las,NA62:2021bji,Esteban:2021tub,Berryman:2022hds,Dev:2024ygx,Chauhan:2024fas,Foroughi-Abari:2025mhj} 
and cosmological observations~\cite{Cyr-Racine:2013jua,Archidiacono:2013dua,Lancaster:2017ksf,Huang:2017egl,Oldengott:2017fhy,Kreisch:2019yzn,Mazumdar:2019tbm,Park:2019ibn,Escudero:2019gvw,Das:2020xke,Mazumdar:2020ibx,Brinckmann:2020bcn,Kreisch:2022zxp,RoyChoudhury:2022rva,Camarena:2023cku,Das:2023npl,An:2023mkf,Camarena:2024daj,Montefalcone:2025ibh,Libanore:2025ack,He:2023oke,He:2025jwp,Whitford:2025dmq,Perez-Castro:2026muj,Fiorillo:2022cdq,Fiorillo:2023cas,Fiorillo:2023ytr,Aloni:2023tff,Allali:2024anb}.

In this work, we assume the scenario in which DM consists entirely of sterile neutrinos produced via oscillations with active neutrinos in the early universe. DM production is enhanced by the presence of neutrino self-interactions, which modify the in-medium properties and scattering rates of active neutrinos, thereby allowing efficient sterile neutrino DM production~\cite{2020PhRvL.124h1802D}. As in the original DW mechanism, the effective mixing angle, which controls DM production, is strongly suppressed at very early times, while particle interaction rates are high. Conversely, unlike in the original DW scenario, at very late times, once active neutrinos fall out of thermal equilibrium with the rest of the thermal plasma, the production can still continue, as long as the active neutrino self-interactions persist. This aspect of the model relaxes the relic DM abundance constraint on the mixing angle and allows for a range of mixing angles to achieve the correct DM abundance at a fixed particle mass, by the appropriate choice of the neutrino self-interaction strength. The final DM abundance for a given sterile neutrino mass thus depends on a combination of active-sterile mixing angle, the self-interaction mediator mass, and the coupling strength; see \figref{fig:s-curv} for an illustration. 

We explore the full range of effects this model has on structure formation. On the one hand, neutrino self-interactions produce scale-dependent modifications to the linear matter power spectrum $P(k)$, as discussed in the previous literature \cite{Kreisch:2019yzn,He:2023oke,Camarena:2023cku,He:2025jwp}. On the other hand, sterile neutrinos suppress structure on small scales, akin to warm DM cosmologies, regardless of the details of their production mechanism, due to a significant free streaming length. The sterile neutrino DM mass, which primarily affects free streaming, is therefore constrained by small-scale structure observations, including the MW satellite galaxy abundance, phase space density derived for dwarf galaxies, Lyman-$\alpha$ (Ly$\alpha$) forest observations, and strong lensing~\cite{Tremaine:1979we,Boyarsky:2008ju,Watson:2011dw,Viel:2013fqw,Baur:2015jsy,Horiuchi:2015qri,Yeche:2017upn,Irsic:2017ixq,DES:2020fxi,Nadler:2021dft,Villasenor:2022aiy,Irsic:2023equ,An:2023mkf}.

To test all structure-related aspects of this scenario with observational data, we implement, for the first time, a self-consistent approach to simultaneously compute the effects of both sterile neutrinos and neutrino self-interactions on $P(k)$. Using this framework, we confront the full model with Ly$\alpha$ forest and cosmic microwave background (CMB) measurements. As a key novelty in this work, we employ constraints derived from two Ly$\alpha$ likelihoods: an effective field theory (EFT) based full-shape likelihood~\cite{Ivanov:2023yla,Ivanov:2024jtl} and a compressed likelihood derived from the PRIYA simulations~\cite{Bird:2023evb,Fernandez:2023grg}. Both these likelihoods are based on measurements of the one-dimensional Ly$\alpha$ flux power spectrum from the Baryon Oscillation Spectroscopic Survey (BOSS) of the Sloan Digital Sky Survey (SDSS) DR14, including data from the extended BOSS (eBOSS) survey~\cite{eBOSS:2018qyj}. We derive the most stringent observational bounds on the active–sterile mixing angle in the context of the full model, surpassing the lower limits previously inferred from DM relic abundance and Big Bang nucleosynthesis (BBN) constraint on mediator mass. Our constraints are also comparable to those obtained from a combination of laboratory based experimental limits on self-interactions and BBN constraint. When combined with independent constraints from X-ray observations~\cite{Boyarsky:2005us,2012JCAP...03..018W,Horiuchi:2013noa,Perez:2016tcq,2020Sci...367.1465D} and MW satellite galaxy abundance~\cite{An:2023mkf}, our constraints can rule out the majority of the parameter space for sterile-neutrino DM production. The key results are presented in \figref{fig:constr_th}.

This paper is organized as follows. In Section~\ref{sec:mod}, we introduce the model for sterile neutrino DM production in the presence of self-interacting neutrinos. In Section~\ref{sec:cosmo}, we describe the impact of this DM model on the matter power spectrum and outline the methodology used for its computation. Section~\ref{sec:data} presents the data sets used and analyses performed. We present our main results for the constraints on the sterile neutrino DM parameters and compare them with the existing bounds in Section~\ref{sec:results}. Finally, Section~\ref{sec:conclusions} summarizes our findings and discusses possible future directions.

\section{Model}\label{sec:mod}
We focus on the sterile neutrino DM production mechanism proposed in Ref.~\cite{2020PhRvL.124h1802D}. This model introduces a new self-interaction among active neutrinos\footnote{Vector mediated nonstandard interaction has also been proposed for efficient sterile neutrino DM production~\cite{Kelly:2020aks,Chichiri:2021wvw}, however, they have been ruled out recently~\cite{Vogel:2025gan}.}, described by the Lagrangian
\begin{equation}
    \mathcal{L} \supset \frac{\lambda_\phi}{2} \nu_a \nu_a \phi\,,
\end{equation}
where $\lambda_\phi$ denotes the strength of coupling to a new complex mediator $\phi$ with mass $m_\phi$, and $a$ corresponds to different lepton flavors: $e, \mu$, and $\tau$. A Fermi-like constant (or the effective coupling) is given by $G_{\rm eff} = \lambda_\phi^2 / m_\phi^2$, and the self-interaction rate through the new mediator is $\Gamma_{\phi\nu_a\nu_a} \propto G_{\rm eff}^2$ for a heavy mediator; for a light mediator, there is a more complex dependence on $\lambda_\phi$ and $m_\phi$, as described in Ref.~\cite{2020PhRvL.124h1802D}. This self-interaction allows for the efficient production of sterile neutrinos, so they can account for all of the DM in the present-day universe. In this scenario, DM can be produced via three distinct production channels, denoted in the original work as Case A (production through a heavy-mediator self-interaction), Case B (on-shell decay of a mediator with a suppressed in-medium mixing angle), and Case C (on-shell decay of a mediator with a mixing angle close to its vacuum value)~\cite{2020PhRvL.124h1802D,An:2023mkf}.

To track the abundance of DM, we use the modified Boltzmann equation at fixed energy ~\cite{1994PhRvL..72...17D,Abazajian:2005gj,Hansen:2017rxr,2020PhRvL.124h1802D,An:2023mkf},
\begin{equation}\label{eq:Boltzmann}
\frac{\mathrm{d} f_{\nu_s}}{\mathrm{d}{z}}  = \frac{\Gamma_\mathrm{tot}}{4H z} \sin^22\theta_\mathrm{eff} f_{\nu_a}\,,
\end{equation}
where $f_{\nu_s}$ and $f_{\nu_a}$ are the phase-space distribution (PSD) functions for the sterile neutrino and active neutrinos, respectively. Here, $H$ is the Hubble parameter; $\Gamma_\mathrm{tot}$ is the total interaction rate of active neutrinos, including both weak and self-interaction contributions; $z$ is cosmological redshift; $T_\gamma$ and $T_{\nu_a}$ represent the photon temperature and active neutrino temperature, respectively, and they are equal at early times. Finally, $\theta_{\rm eff}$ denotes the effective mixing angle within the hot plasma, given by
\begin{equation}\label{eq:theff}
    \sin^22\theta_\mathrm{eff} \simeq \frac{\Delta^2 \sin^22\theta}{\Delta^2 \sin^22\theta + \Gamma^2/4 + (\Delta \cos2\theta - V_T)^2 }\,,
\end{equation}
where $\theta$ represents its vacuum value; $\Delta = m_4^2/(2E)$ is the vacuum oscillation frequency; $E$ is the neutrino energy; and $V_T$ is the thermal potential of active neutrinos. 

In this framework, DM production rate $\Gamma_\mathrm{prod.}=\frac{1}{2}{\Gamma_\mathrm{tot}}\sin^22\theta_\mathrm{eff}$ is generally only significant within a finite temperature range---while both factors in this expression have a non-vanishing value---and it asymptotes to zero otherwise. Namely, at early times, the effective mixing angle given by \eqnref{eq:theff} vanishes, as the interaction rates and the neutrino potential are both very high $\Gamma, V_T \gg \Delta$. 
Conversely, as the temperature drops at late times, $\Gamma, V_T \to 0$ and the effective mixing angle defaults to its vacuum value; at this time, the production stops since $\Gamma_\mathrm{tot}$ vanishes. Therefore, efficient production occurs only at a specific time when appropriate conditions are met. This framework accurately captures the modified thermal history of the early Universe and its consequences for non-standard sterile neutrino distribution functions. Please refer to Ref.~\cite{2020PhRvL.124h1802D} for more details.

To track DM production, we first integrate \eqnref{eq:Boltzmann} over the redshift range where DM production is most efficient while DM is still ultra-relativistic, and obtain the exact non-thermal $f_{\nu_s}$. The present relic abundance of the sterile neutrino DM is then calculated as
\begin{equation}\label{eq:relic}
    \Omega_c = \left( \frac{2891.2}{1.05 \times 10^{-5} h^2} \right)\, \frac{n_{\nu_s} m_4}{s}\, \mathrm{GeV}^{-1}\, 
\end{equation}
where $n_{\nu_s}$ is the number density of sterile neutrinos, $s$ is the entropy density of the universe and $h$ is the reduced Hubble parameter. This numerically computed non-thermal PSD for sterile neutrino DM is then passed directly into our modified version of \texttt{CLASS} to compute the linear matter power spectrum $P(k)$. Further details regarding the exact procedure employed to compute P(k) within our model are provided in the following section.

\begin{figure}[!t]
\centering
\includegraphics[width=0.5\textwidth]{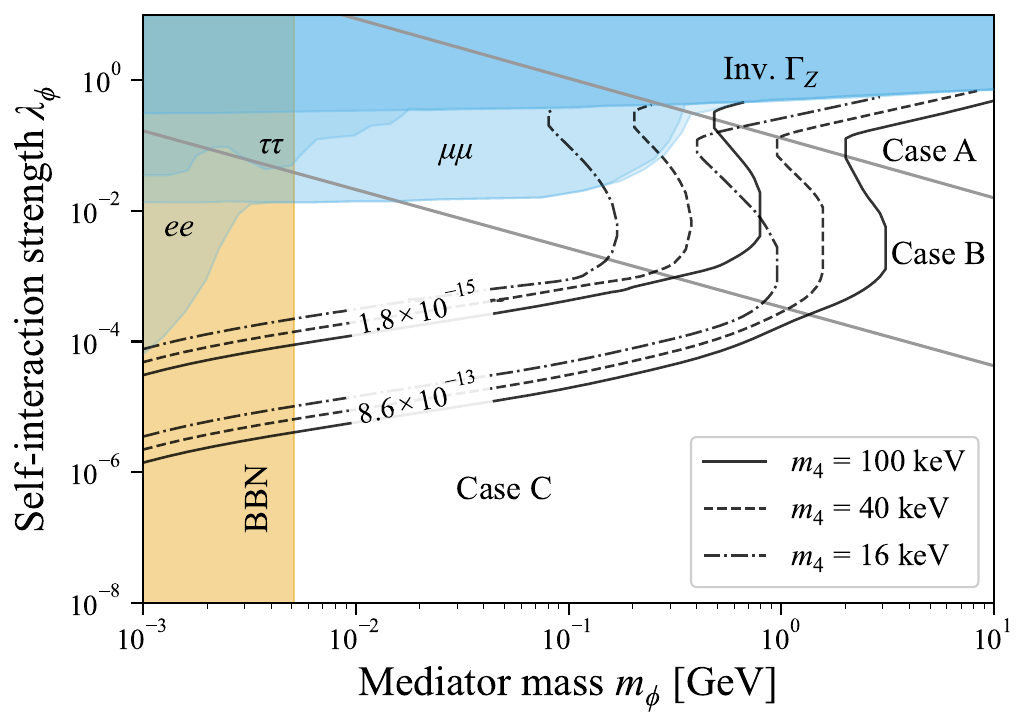}
\caption{Self-interaction strength $\lambda_\phi$ required to produce the sterile neutrino DM consistent with observed DM relic density, $\Omega_c h^2 = 0.12$~\cite{Planck:2018vyg}, as a function of mediator mass, for several sterile-neutrino masses and mixing angles (listed in the legend). The characteristic $\mathcal{S}$-shaped curves shown by solid, dashed, and dot-dashed lines correspond to particle masses of $m_4= 100$, $40$, and $16$ keV, respectively; the curves are shown for two different values of $\sin^2(2\theta)$, as labeled in each set of curves. The gray solid lines separate the parameters space corresponding to the three production channels: Cases A, B, and C, as discussed in the text. The blue shaded regions are excluded by neutrino experiments involving the invisible $Z$ decay (Inv. $\Gamma_Z$) and flavor-dependent constraints ($ee$, $\mu\mu$, and $\tau\tau$ denote the type of flavor-specific coupling, corresponding to $\nu_e$, $\nu_\mu$, and $\nu_\tau$, respectively)~\cite{2018PhRvD..97g5030B,Blinov:2019gcj,Berryman:2022hds,Esteban:2021tub}. The orange shaded region is inconsistent with BBN and the light-element-abundance measurements~\cite{Huang:2017egl,Blinov:2019gcj,Escudero:2019gvw,2020PhRvL.124h1802D}.}\label{fig:s-curv}
\end{figure}
In this DM model, the abundance of DM relics depends on four parameters: $m_4$, $\theta$, $\lambda_\phi$, and $m_\phi$. \figref{fig:s-curv} illustrates the characteristic $\mathcal{S}$-shaped curve that captures the relationship between $\lambda_\phi$ and $m_\phi$, which is obtained by requiring that sterile neutrino DM is produced at a fixed DM abundance, $\Omega_c h^2 = 0.12$~\cite{Planck:2018vyg}. We show this curve for a combination of three sterile neutrino masses and two mixing angles as labeled. Note that the values of $m_4$, $\theta$, and $m_\phi$ are chosen independently of each other, and the requirement to obtain the correct relic abundance is satisfied by computing the interactions strength $\lambda_\phi$ as a function of these three parameters. Different parts of the curves correspond to the three production channels discussed previously; for clarity, we draw the two diagonal gray lines to separate these regions: the region between the diagonal lines corresponds to Case B, the region below the lower diagonal line corresponds to case C, and the region above the upper line to Case A.

We note that, as expected, the increase in the mixing angle generally leads to a lower value of $\lambda_\phi$ (for a given mediator and sterile-neutrino mass) to maintain a fixed production rate and provide the required abundance of DM. Similarly, given the interaction rate scaling, higher values of $\lambda_\phi$ generally require higher values of $m_\phi$.\footnote{A notable exception is Case B, where the interactions decouple while the mixing angle is still suppressed. In this case, the suppression of the mixing angle and the mediator mass pose competing effects that leads to the opposite proportionality between $\lambda_\phi$ and $m_\phi$; see Ref.~\cite{2020PhRvL.124h1802D} for details.} We also show the portion of the parameter space excluded by various laboratory experiments as blue shaded region in the Figure~\cite{Barger:1981vd,ALEPH:2005ab,2018PhRvD..97g5030B,Agostini:2015nwa,Blum:2018ljv,Berryman:2018ogk,Blinov:2019gcj,Brdar:2020nbj,Lyu:2020lps,PIENU:2020las,NA62:2021bji,Esteban:2021tub,Berryman:2022hds,Dev:2024ygx,Foroughi-Abari:2025mhj}. The laboratory bounds include constraints from the decay width of invisible z-decay~\cite{Blinov:2019gcj,Brdar:2020nbj,Foroughi-Abari:2025mhj} and flavor-dependent constraints from observations of neutrino-less double $\beta$ decay~\cite{Agostini:2015nwa,Blum:2018ljv}, meson decays~\cite{Barger:1981vd,Berryman:2018ogk,Blinov:2019gcj,PIENU:2020las,NA62:2021bji,Dev:2024ygx} and $\tau$-decay~\cite{Blinov:2019gcj,Brdar:2020nbj}. The orange shaded region is excluded by the BBN and the measurements of the light element abundances~\cite{Huang:2017egl,Blinov:2019gcj,Escudero:2019gvw,2020PhRvL.124h1802D}.

\section{Matter Power Spectrum}\label{sec:cosmo}

\begin{figure}[!tbp]
\centering
\includegraphics[width=0.5\textwidth]{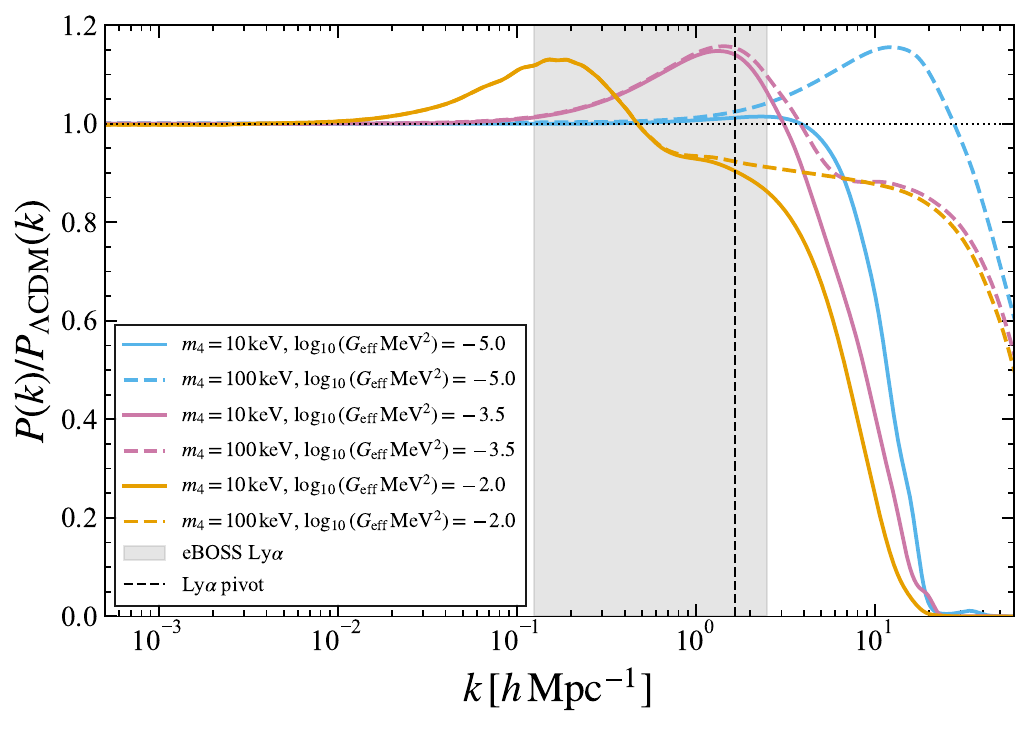}
\caption{Ratio of the linear matter power spectrum in a cosmological model with sterile-neutrino DM and self-interacting neutrinos to the standard CDM case is shown for a range of DM particle masses $m_4$ and effective couplings $G_\mathrm{eff}$, shown in the legend. The power spectra exhibit two key features: a bump, which depends on effective self-interaction coupling constant, and a sharp cutoff, related to the mass of sterile neutrinos. In each case, other model parameters are set such that neutrino DM has the observed DM relic abundance. The gray shaded region represents the scales probed by Ly$\alpha$ forest observations from the eBOSS survey~\cite{eBOSS:2018qyj,He:2023oke,He:2025jwp}. The black dashed line denotes the pivot scale at which the reduced likelihood from the PRIYA simulation is evaluated~\cite{Bird:2023evb,Fernandez:2023grg,He:2023oke,He:2025jwp}.}\label{fig:pk}
\end{figure}

The key observable on which we focus is the linear matter power spectrum $P(k)$. In this work, for the first time, we compute the matter power spectrum simultaneously accounting for two effects: i) small-scale suppression of power resulting from sterile-neutrino DM free streaming and ii) scale-dependent enhancement and suppression of power at a \textit{separate scale}, resulting from active neutrino self-interactions and their delayed decoupling. Both effects are illustrated in \figref{fig:pk}. 

To compute the matter power spectrum $P(k)$ within this framework, we start with the implementation of the modified Boltzmann hierarchy equations for interacting active neutrinos from Ref.~\cite{Kreisch:2019yzn}, into the publicly available CLASS code~\cite{2011arXiv1104.2932L,class2011,He:2023oke,He:2025jwp}. We extend this code to include the model parameters $\Theta\in\{m_4, m_\phi, \lambda_\phi, \theta\}$ consistently. We note that the small-scale power cutoff is primarily controlled by the free streaming of sterile-neutrino DM, while the excess power primarily depends on the self-interaction strength; however, the two effects are not entirely independent of each other, as we can constrain $\Theta$ to always result in the given relic abundance of DM $\Omega_{c}$. To ensure that this consistency relation is satisfied in our version of CLASS, we first determine the mixing angle required to produce a given $\Omega_{c}$ for any given set of \{$m_4$, $m_\phi$, $\lambda_\phi$\}. To achieve this within reasonable computational time, we build a neural network based emulator that predicts the correct mixing angle for the given values of $m_\phi$, $\lambda_\phi$, $m_4$, and $\Omega_c$, trained on a large set of exact solutions obtained by directly solving \eqnref{eq:relic}. The emulator is then integrated as a part of our CLASS code, and is described in detail in an Appendix~\ref{sec:AppA}. The next step involves the computation of (non-thermal) PSD of sterile neutrino for the resulting parameters and then feed it to the modified CLASS code to compute the matter power spectrum. We have consistently incorporated all these steps to build a robust pipeline for the computation of $P(k)$ within our sterile neutrino DM model. We explicitly emphasize that the neural-network based emulator is used exclusively to compute the mixing angle required to produce a provided DM density for given $m_4, m_\phi$, and  $\lambda_\phi$ values; it does not approximate or interpolate the PSD itself, the PSD is obtained by solving the modified Boltzmann equation, accounting for all physical effects and the modified thermal history in the early Universe, as described in Sec.~\ref{sec:mod}.

In \figref{fig:pk}, we show the resulting $P(k)$ (normalized to the CDM case), for few reference sterile neutrino masses $m_4 = 10$ and $100$ keV, and effective coupling $\log_{10} (G_{\rm eff}\,{\rm MeV}^{2}) = -2.0, -3.5,$ and $-5.0$. We see that as the value of $G_{\rm eff}$ increases, the excess power (the ``bump'') feature that corresponds to the horizon scale at neutrino decoupling moves to smaller wavenumber $k$ (decoupling at later times), whereas increasing the sterile neutrino mass shifts the cutoff to larger $k$ (smaller length scales corresponding to a smaller free streaming length), as expected.

\section{Data and Analysis}\label{sec:data}

In this work, we leverage Ly$\alpha$ forest observations in combination with Planck CMB data~\cite{Planck:2019nip} and CMB lensing data from the Atacama Cosmology Telescope
(ACT) DR6 and Planck PR4 NPIPE~\cite{ACT:2023kun,ACT:2023dou,Carron:2022eyg} to assess the viable parameter space for this model. For Ly$\alpha$, we employ two likelihoods derived using two different forward-modeling techniques, both based on the 1D Ly$\alpha$ flux power spectrum data from SDSS DR14 BOSS and eBOSS~\cite{eBOSS:2018qyj} quasars. The first model is based on the EFT of large scale structure, and calibrated on Sherwood simulations~\cite{Bolton:2016bfs,Givans:2022qgb}, with the associated likelihood based on eBOSS data from the redshift range $3.2 < z < 4.2$ \cite{Ivanov:2023yla,Ivanov:2024jtl}; we label this likelihood as ``Ly$\alpha$--EFT'' in our figures. The second model is based on the PRIYA simulation suite~\cite{Bird:2023evb,Fernandez:2023grg}, which provides a compressed likelihood in the form of Gaussian priors on the amplitude $\Delta_\star^2 = \frac{k_\star^3 P(k_\star, z_\star)}{2\pi^2}$ and slope $n_\star = \left. \frac{\mathrm{d}\log P(k, z)}{\mathrm{d}\log k} \right|_{k_\star, z_\star}$\footnote{These two quantities are denoted by $\Delta_L^2$ and $n_{\rm eff}$ in Ref.~\cite{He:2025jwp}.}, evaluated at the pivot scale $k_\star = 0.009 $ s/km $\sim 1$ Mpc$^{-1}$ and pivot redshift $z_\star=3$~\cite{He:2025jwp}; we label this case as ``Ly$\alpha$--PRIYA''.

As explained above, our model introduces four parameters ($m_4, m_\phi, \lambda_\phi, \theta$) in addition to the six standard cosmological parameters into the analysis. However, the requirement that sterile neutrinos are produced at the correct abundance $\Omega_\mathrm{c}$ poses a constraint given by \eqnref{eq:relic}, which we use to fix the mixing angle. Furthermore, $m_\phi$ and $\lambda_\phi$ combine to set the effective coupling $G_{\rm eff}$, leaving $G_{\rm eff}$, $m_4$ and the six standard cosmological parameters as the only independent parameters of this cosmological model. As described in Sec.~\ref{sec:mod}, $m_4$ controls the small-scale free-streaming cutoff and $G_{\rm eff}$ governs the location of the ``bump'' feature in $P(k)$. In this work, we consider the range of neutrino sterile masses $m_4 \in [10, 1000]\, {\rm keV}$. In this mass range, modifications to the linear matter power spectrum due to sterile neutrino free streaming within the range of scales ($0.1$ hMpc$^{-1}$ $\lessapprox k < 2.5 $ hMpc$^{-1}$)  probed by Ly$\alpha$ forest observations from the eBOSS survey are negligible, and the onset of sharp suppression in $P(k)$ occurs at even higher $k$ scales (see Fig.~\ref{fig:pk}). Since the observable modifications to $P(k)$ within the Ly$\alpha$ forest probed scales are mainly dominated by the active neutrino self-interaction dynamics rather than sterile neutrinos, the observational likelihood remains primarily sensitive to the exact physical effects constrained in pure self-interacting neutrino frameworks. Hence, rather than re-performing the full Markov Chain Monte Carlo (MCMC) analysis, we reliably extract constraints by applying a post-processing parameter transformation to the MCMC chains generated by Ref.~\cite{He:2025jwp}, which analyzed self-interacting neutrino cosmology using these same data sets. Such transformations are a standard and widely accepted approach in cosmological inference, conceptually identical to deriving posteriors for parameters like $S_8$ from primary chains of $A_s$ and $\Omega_m$, and do not require re-running full sampling chains provided the underlying posterior distributions are well-sampled. We adopt this approach in our analysis due to severe computational limitations, as performing a full MCMC exploration with neutrino self-interactions alongside a non-thermal sterile neutrino PSD inside \texttt{CLASS} is computationally expensive. Furthermore, while active neutrino self-interaction controls DM production, thereby coupling $G_{\rm eff}$ and $\Omega_c$, the observational data itself do not produce any significant correlation between these two parameters, as demonstrated in Figs.~$6$ and $7$ of Ref.~\cite{He:2025jwp}. This empirical lack of correlation justifies treating the posterior in the $(G_{\rm eff}, \Omega_c)$ plane as an approximately factorized distribution during our parameter mapping step, safely translating these posteriors into bounds on the parameters of this model.

To translate the MCMC chain into constraints on the parameters of this model, we first randomly sample $m_\phi \in [10^{-3}, 10]\, {\rm GeV}$ and $\lambda_\phi \in [10^{-9}, 1]$ retaining only those combinations that are consistent with the values of $G_{\rm eff}$ given in the original MCMC chains.  Because the relevant cosmological observables depend on $\l_\phi$ and $m_\phi$ exclusively through the effective coupling $G_{\rm eff} = \lambda_\phi^2/m_\phi^2$, the model exhibits an inherent degeneracy between $\lambda_\phi$ and $m_\phi$. Our sampling procedure explicitly preserves this degeneracy by exploring the full parameter ranges of $m_\phi$ and $\lambda_\phi$ considered in this work along directions of constant $G_{\rm eff}$. For a given set of sterile neutrino mass $m_4 \in [10, 1000]$ keV, these selected values of $m_\phi$ and $\lambda_\phi$, together with the corresponding value of $\Omega_c$ from the MCMC chains, are then passed to our emulator discussed in the previous section.  The emulator returns the mixing angle required to reproduce the correct relic density. In this way, we construct derived MCMC chains for the parameters of our model. Crucially, because our pipeline scans over the entire parameter volume, covering all production channels simultaneously, rather than isolating individual production regimes, the resulting derived MCMC chains encapsulate the multi-modal nature of the parameter space. The resulting constraints are presented and discussed in detail in the next section.

\section{Results}\label{sec:results}

\begin{figure}[t]
\centering
\includegraphics[width=0.5\textwidth]{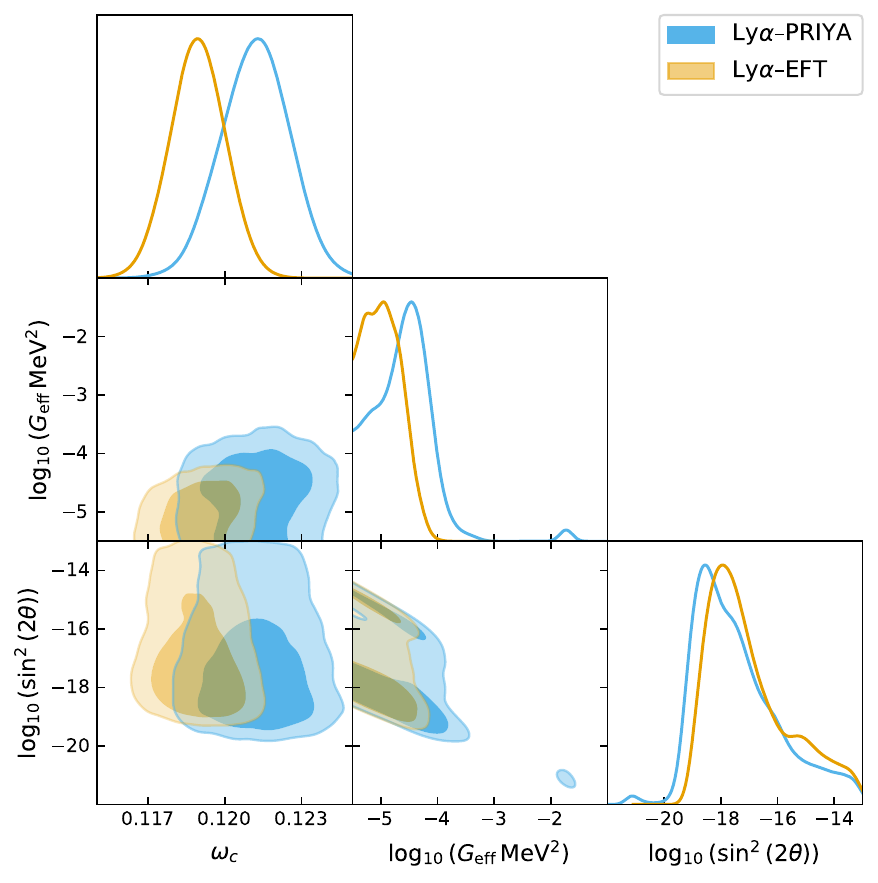}
\caption{1-D and 2-D marginalized posterior probability distributions for the active-sterile mixing angle, DM relic density, and the effective coupling constant between active neutrinos, assuming a sterile neutrino mass of $m_4 = 100$ keV. The two colors correspond to the results obtained using two different likelihoods based on eBOSS data: PRIYA-simulation and EFT-based likelihoods. The dark and light shaded regions denote the $68\%$ and $95\%$ C.~L.~contours, respectively.}\label{fig:tri_eft}
\end{figure}
\begin{figure}[!htbp]
\centering
\includegraphics[width=0.5\textwidth]{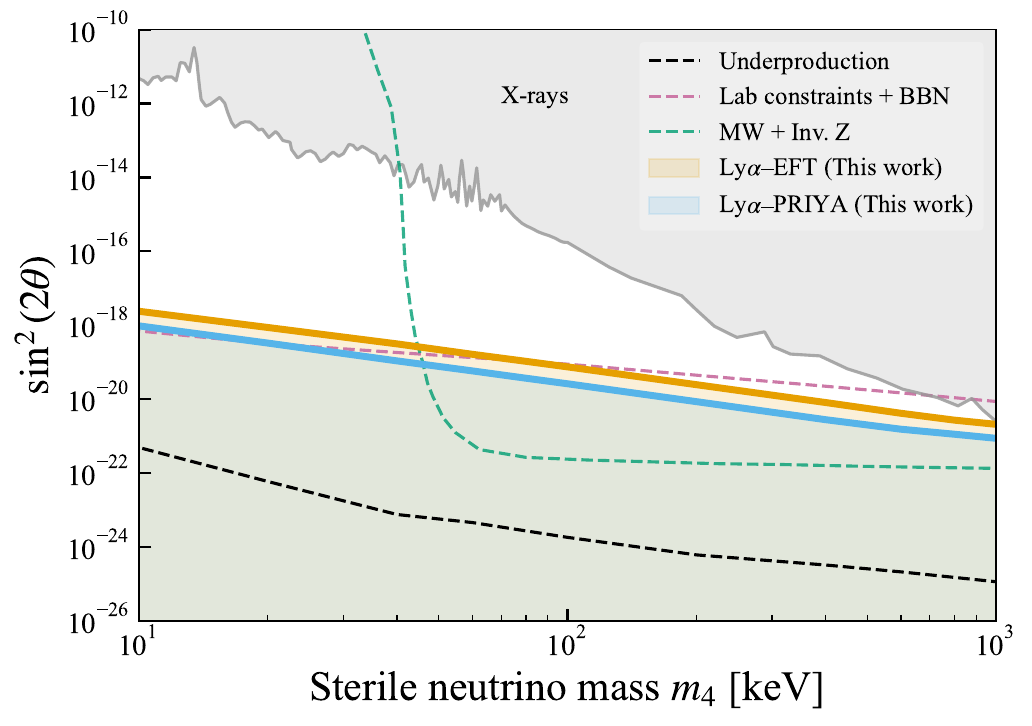}
\caption{Constraints on sterile-neutrino DM from our analysis with Ly$\alpha$ forest observations. The two results of the current work are shown as the $95\%$ C.~L.~lower bounds on the mixing angle between active and sterile neutrino, as a function of sterile-neutrino mass, obtained from the EFT and on PRIYA models and likelihoods, respectively. The thickness of the colored constraint lines represent the propagated $\sim 7\%$ (95th percentile) emulator uncertainty on our constraints. We also show the lower bound on the mixing angle from a combine analysis of BBN  data \cite{Huang:2017egl,Blinov:2019gcj,Escudero:2019gvw,2020PhRvL.124h1802D} with laboratory bounds on neutrino self-interactions~\cite{2018PhRvD..97g5030B,Blinov:2019gcj,Berryman:2022hds,Esteban:2021tub} (pink dashed line). The combined bound from MW satellite abundance and invisible $Z$ decay from Ref.~~\cite{An:2023mkf} is shown as the dashed green line; this bound is only valid under conditions discussed in the text. The black dashed line is lower bound on the mixing angle required to produce the observed DM abundance as measured by \textit{Planck} \cite{Planck:2018vyg}. The shaded gray region is ruled out by X-ray observations~\cite{Boyarsky:2005us,2012JCAP...03..018W,Horiuchi:2013noa,Perez:2016tcq,2020Sci...367.1465D}. Overall, we note that the results of this work provide the most stringent observational bound in this parameter space, surpassing the underproduction limits from the CMB by several orders of magnitude. Note that the result is comparable to the most stringent laboratory constraints in the full sterile-neutrino mass range.}\label{fig:constr_th}
\end{figure}
\begin{figure}[!htbp]
\centering
\includegraphics[width=0.5\textwidth]{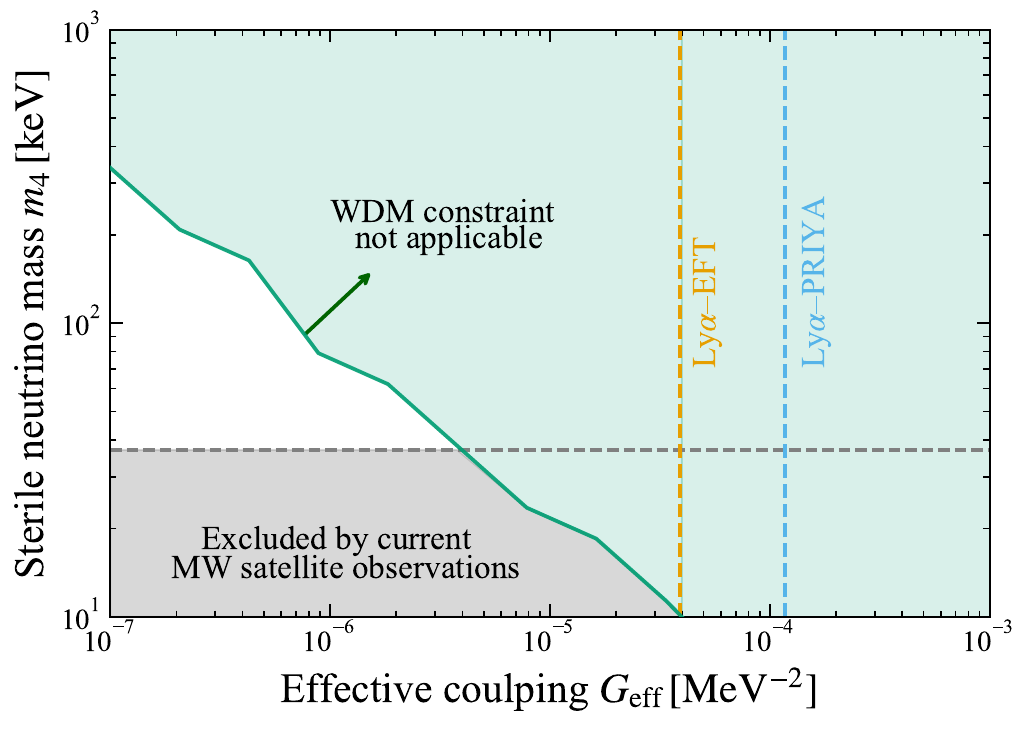}
\caption{Schematic depiction of the $m_4$--$G_{\rm eff}$ parameter space where constraints on sterile-neutrino DM obtained while neglecting self-interaction effects on the matter power spectrum (``bump feature'') are not applicable (green shaded region). The blue and orange dashed lines indicate the upper limit on the $G_{\rm eff}$ (at $95\%$ C.~L.~) from the Ly$\alpha$ forest analysis, and the gray dashed line represent the lower bound on the sterile neutrino DM obtained from MW satellite abundances utilizing WDM simulations~\cite{An:2023mkf}. This bound can be applied to our model only where the effective coupling strength lies outside the green shaded region; within the shaded region, the power spectrum contains a feature that is not strictly modeled in the MW analysis. Gray shaded region represent the  parameters space excluded by these MW bounds for our model.}\label{fig:constr_val}
\end{figure}

\figref{fig:tri_eft} shows the 1-D and 2-D marginalized posteriors for $\omega_c= \Omega_ch^2$, $\log_{10}(G_{\rm eff}~{\rm MeV^{2}})$, and $\log_{10}(\sin^2({2\theta}))$ at $68 \%$ and $95\%$  confidence level (C.~L.~), for Ly$\alpha$--PRIYA (blue color) and Ly$\alpha$--EFT (orange color) likelihoods, for sterile neutrino mass of $m_4 = 100$ keV. We repeat the analysis for a set of sterile neutrino mass values in the range $10$--$1000$ keV, and obtain the lower limit on the mixing angle for each mass in \figref{fig:constr_th}. The two solid colored lines in that figure correspond to the two likelihoods considered in this work, and the shaded regions are excluded at $95\%$ C.~L. The thickness of these constraint lines captures the uncertainty introduced by emulator in our predictions. It should be noted that our emulator has $\sim 7\%$ error at 95th percentile of the test error distribution.
It is evident from Fig.~\ref{fig:constr_th} that the uncertainty introduced by the emulator on our constraints is significantly smaller than the total width of the excluded parameter regions and remains highly subdominant.

We note a few features of posterior probability distributions in \figref{fig:tri_eft}. First, since $G_{\rm eff}=\lambda_\phi^2/m_\phi^2$, the interaction strength and mediator mass are almost perfectly degenerate and cannot be constrained independently, so we omit them from the plots. 
We further see in \figref{fig:tri_eft} that $\omega_c$ values inferred from the two likelihoods are slightly different; this parameter is mainly constrained from CMB data, and the slight difference may be due to the slightly different $\sigma_8$ values preferred by the two Ly$\alpha$ likelihoods, and the degeneracy between $\omega_c$ and $\sigma_8$; see Ref.~\cite{He:2025jwp} for more details.

In addition, the posterior probability distribution for the mixing angle shows two modes, observed in \figref{fig:tri_eft}.  This feature flows directly from the non-trivial relationship between $\lambda_\phi$ and $m_\phi$ shown in \figref{fig:s-curv}: each mode corresponds to the dominance of a particular production channel for sterile neutrinos. We emphasize that this feature is an intrinsic physical property of the underlying DM production mechanism rather than a data-driven cosmological degeneracy or sampling artifact; the observational likelihood depends on these parameters ($\lambda_\phi$ and $m_\phi$) through $G_{\rm eff}$. Consequently, this multi-valued mapping does not compromise or invalidate our reparameterization procedure. Hence, the DM abundance no longer exhibits a one-to-one relationship between $\theta$ and $\Omega_c$. Instead, for a given $m_4$ and $G_{\rm eff}$ consistent with observations, the correct DM can be produced through distinct channels corresponding to different values of $\theta$. In particular, the lower-$\theta$-value mode in the posterior is primarily associated with production through Cases B and C, while the higher $\theta$ value mode corresponds primarily to production through Case A. Since most of the higher $\theta$ values will be excluded by X-ray observations, Case A is largely ruled out, while the production through the C and B channels is still viable. This observation may provide further guidance for future model building in light of observational constraints. Finally, we also emphasize that our neural-network emulator, trained across the entire multi-dimensional parameter space, quite accurately captures these distinct production branches, ensuring that the mapped posteriors self-consistently preserve this physical multimodality.

Finally, there is no upper bound on the mixing angle from this analysis---the upper limits seen in the triangle plots are an artifact resulting from the lack of posterior samples in that region of the parameter space, and the bound is \textit{not} physical. However, the lower bound on the mixing angle is well sampled and robust. 

Our main results are presented in \figref{fig:constr_th}. For comparison with our current constraints, this figure also shows the lower limit on the mixing angle derived from laboratory experiments \cite{2018PhRvD..97g5030B,Blinov:2019gcj,Berryman:2022hds,Esteban:2021tub} and from primordial light element abundances \cite{Huang:2017egl,Blinov:2019gcj,Escudero:2019gvw,2020PhRvL.124h1802D}, shown in the pink dashed line and discussed in Section~\ref{sec:mod}. We further show the upper bound on mixing angles from X-ray observations~\cite{Boyarsky:2005us,2012JCAP...03..018W,Horiuchi:2013noa,Perez:2016tcq,2020Sci...367.1465D} (gray shaded region), and the lower bound of the same parameter from DM abundance measurements\footnote{\textit{Planck} measurement of DM relic abundance~\cite{Planck:2018vyg}, combined with light element abundance measurements~\cite{Huang:2017egl,Blinov:2019gcj,Escudero:2019gvw,2020PhRvL.124h1802D} implies an ``underproduction'' bound by imposing a lower limit for the mixing angle, for $\lambda_\phi\leq 1$~\cite{2020PhRvL.124h1802D}.}. The viable parameter space is represented by the white color in this Figure. 

Finally, we include in the same figure the combined constraint on sterile neutrinos from the MW satellite galaxy abundance and invisible z-decay as the green dashed line ~\cite{An:2023mkf}. The MW satellite constraints were obtained using the warm DM (WDM) simulations.  However, we note that the analysis in Ref.~\cite{An:2023mkf} did \textit{not} model the effects of neutrino self-interactions on $P(k)$ and has only focused on the WDM-like cutoff at small scales. Therefore, this constraint strictly applies only to a portion of our parameter space where the bump feature induced by neutrino self-interaction is removed by the cutoff caused by DM free streaming. We show a schematic depiction of the applicability realm of the MW bound in our model in \figref{fig:constr_val}. For values of $m_4$ and $G_{\rm eff}$ within the green shaded region, labeled "WDM constraint not applicable", in the same figure, a modeling of the MW satellite population in models with self-interactions is necessary to fully understand the bound on the sterile-neutrino mass. Hence, only the gray shaded part of $m_4$-$G_{\rm eff}$ parameter space would be ruled out by the current MW constraint.

Overall, \figref{fig:constr_th} demonstrates that the Ly$\alpha$ forest analyses presented in this work provide the most stringent observational constraint on sterile-neutrino DM produced through active-neutrino self-interactions. This constraint surpasses the underproduction bounds from the DM relic density by several orders of magnitude, and is comparable to the bound obtained by combining the most stringent flavor-dependent laboratory constraints.

\section{Conclusions and discussion}\label{sec:conclusions}
This work explored a specific model of sterile neutrino DM, where the production of the sterile neutrino is facilitated through a new self-interaction among active neutrinos~\cite{2020PhRvL.124h1802D}---a scenario well-motivated from the standpoint of neutrino mass production mechanisms. This model produces unique features in the matter power spectrum, testable by observational data. To understand the observational constraints on the model, we self-consistently include the effect of both the delay in active-neutrino free streaming due to a new self-interaction, and the effect of sterile-neutrino free streaming, in the computation of the linear matter power spectrum. 

We then compare this model with Ly$\alpha$ forest observations. Specifically, we employ two Ly$\alpha$ likelihoods that measure the amplitude and shape of the linear matter power spectrum on small scales ($k\approx0.1$--$2$ $h {\rm Mpc}^{-1}$): an EFT-based full shaped likelihood~\cite{Ivanov:2023yla,Ivanov:2024jtl}; and a likelihood based on PRIYA simulation~\cite{Bird:2023evb,Fernandez:2023grg}. Our analysis results in the strongest observational constraint on the sterile neutrino DM production to-date, comparable to the bounds obtained from flavor-dependent laboratory experiments~\cite{2018PhRvD..97g5030B,Blinov:2019gcj,Berryman:2022hds,Esteban:2021tub}; \figref{fig:constr_th} shows our key result. When considered in conjunction with constraints from X-ray observations~\cite{Boyarsky:2005us,2012JCAP...03..018W,Horiuchi:2013noa,Perez:2016tcq,2020Sci...367.1465D} and MW satellite abundance~\cite{An:2023mkf}, a large portion of the viable parameter space for this model is ruled out.

Other observational bounds on the mass of sterile neutrino DM also exist, as discussed in Secs.~\ref{sec:intro} and \ref{sec:results}. We note that the strongest lower bounds on the sterile neutrino mass originate from measurements of small-scale structure, such as the MW satellite galaxy population \cite{An:2023mkf}, and the high-resolution Ly$\alpha$ forest measurements from BOSS and XQ-100 surveys \cite{Baur:2015jsy,Yeche:2017upn}.
However, these previous studies model cosmologies with strict suppression of power at large $k$, where the power spectrum resembles that of warm DM; as a result, their bounds are \textit{not} readily applicable to cosmologies with a more complex $k$--dependence, as the one we focus on in this work. The derivation of the mass bounds from these observables requires dedicated forward modeling of structure in cosmologies with both neutrino self-interaction and sterile neutrino free streaming and is enabled by the power spectrum calculation presented in this work. We leave a dedicated derivation of the new sterile-neutrino mass bound from high-resolution Ly$\alpha$ data to future work. 

We further note a few potential caveats of our present analysis and outline directions for future work. Importantly, the compressed likelihood inferred from the PRIYA simulation suite relies on simulations which varied the slope and amplitude of the matter power spectrum, without explicitly modeling self-interacting neutrinos. However, the current observed 1-D Ly$\alpha$ flux power spectra are primarily sensitive to a range of scales in $P(k)$ that are typically well-described by local shifts in amplitude and tilt within the observational range. It is generally only in models with highly oscillatory features in the matter power spectrum that this approximation breaks down; the model we explore does not exhibit such oscillations and is well within the space of modifications probed by the simulations. Ref.~\cite{He:2025jwp} explicitly tested the validity of approximating the effect of self-interacting neutrinos with a slope and amplitude shift. For their best-fit parameters with $\log_{10}(G_{\rm eff}~{\rm MeV^{2}})\sim -5$, the approximation was good at $<1\%$ over the redshift and scale ranges probed by the eBOSS Ly$\alpha$ forest~\cite{He:2025jwp}. The approximation is likely less accurate for the larger values of $\log_{10}(G_{\rm eff}~{\rm MeV^{2}})$ shown in \figref{fig:pk}, but these values are strongly disfavored by the data. Our use of the PRIYA reduced likelihood is thus justified for our current purposes but should be re-assessed in future data analyses. An exact treatment of this model across the full parameter space will ultimately require dedicated hydrodynamical Ly$\alpha$-forest simulations incorporating the effects of active-sterile neutrino dynamics, which is beyond the scope of this work.

Additionally, we also perform an independent analysis with full-shape Ly$\alpha$ likelihood based on EFT. The EFT-based likelihood does not assume a particular shape for the matter power spectrum but uses a deterministic  cosmology-independent relationship between the non-linear EFT parameters and the linear bias coefficient (motivated by the clustering of dark matter halos).  These functions are calibrated on Sherwood and ACCEL2 cosmological simulations using a Planck-based cosmology~\cite{Ivanov:2024jtl,deBelsunce:2024rvv, deBelsunce:2025bqc}.  In addition, the EFT-based likelihood also takes a different approach in modeling effects like patchy helium reionization and thus returns results that differ somewhat from PRIYA-based analyses~\cite{Ivanov:2024jtl,deBelsunce:2024rvv}. The consistency between the results obtained using these two conceptually distinct Ly$\alpha$ likelihood demonstrates that our final limits are robust and are not artifacts of any model mismatch within the likelihood compression pipeline. A detailed comparison between these two approaches is an ongoing research topic, and we leave detailed comparison of their results in the context of a specific cosmological model for future work.

Finally, we highlight that the Ly$\alpha$ forest cosmology is advancing at a rapid pace. Notably, DESI-based likelihoods are becoming available~\cite{Chaves-Montero:2026hqd}, with substantially improved systematic error control, as well as improved statistics. 
Furthermore, there are a range of observations available which probe even smaller length scales, constraining the existence of a cut-off in the matter power spectrum, as well as IGM thermodynamics \cite[e.g.~]{Irsic:2017ixq, Rogers:2020ltq, Rogers:2021byl,Irsic:2023equ, Ho:2025ajo,Ivanov:2025pbu}. With several ongoing and forthcoming spectroscopic programs offering enhanced sensitivity and statistical power, Ly$\alpha$ forest cosmology promises increasingly stringent probes to the DM and beyond standard model physics scenarios.

\begin{acknowledgments} 
VG acknowledges the support from the National Science Foundation (NSF) CAREER Grant No. PHY-2239205, from the Research Corporation for Science Advancement under the Cottrell Scholar Program, and the IBM Einstein Fellowship at the Institute for Advanced Study. This publication was made possible through the support of Grant 63667 from the John Templeton Foundation. The opinions expressed in this publication are those of the author(s) and do not necessarily reflect the views of the John Templeton Foundation nor the NSF. YZ is supported by a Subatomic Physics Discovery Grant (individual) from the Natural Sciences and Engineering Research Council of Canada.
\end{acknowledgments}

\providecommand{\href}[2]{#2}\begingroup\raggedright\endgroup


\newpage
\appendix
\section{Neural Network Emulator for Mixing Angle Prediction}\label{sec:AppA}
This appendix describes the architecture, training, and testing of our neural network (NN)–based emulator designed to predict active--sterile mixing angles from four input parameters: $m_4$, $m_\phi$, $\lambda_\phi$, and $\omega_{c}$. We employ the emulator in place of the computationally expensive numerical procedure required to map a given DM relic density to a mixing angle.

\begin{figure*}[!htbp]
	\centering
	\includegraphics[width=1.0\textwidth]{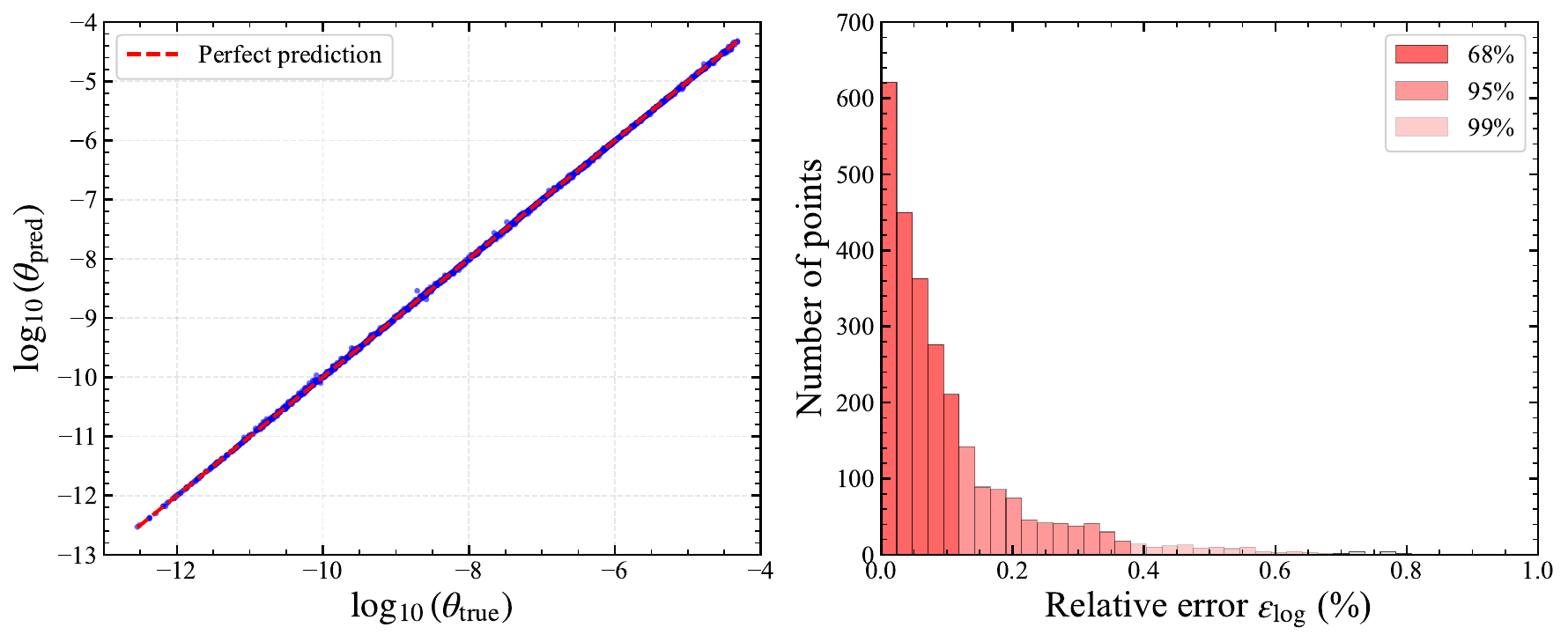}
	\caption{Prediction performance of the NN-based emulator for the active-sterile mixing angle on the independent test dataset. {\bf Left panel:} Predicted values ($\log_{10}(\theta_{\rm pred}$)) plotted against the actual values $\log_{10}(\theta_{\rm true}$)) as shown by the blue points. The red dashed line represent the actual values from the test data. {\bf Right panel:} Distribution of relative percentage errors $\varepsilon_{\rm log}$ for the test data. The red shaded bands indicate regions with $68\%$, $95\%$, and $99\%$ of the total number of test data (from darkest to lightest, respectively).
	}\label{fig:em_th}
\end{figure*}

To build this NN-based emulator, we first generate the data using the full mathematical model. We sample the input parameter space using the Latin Hypercube Sampling (LHS) technique in the ranges $m_\phi \in [10^{-3}, 10] {\rm GeV}, m_4 \in [1, 1000]\, {\rm keV},  \sin^2(2\theta) \in [10^{-25}, 10^{-6}]$ and $\omega_{c} \in [0.05, 0.16]$. For each sample point, we numerically solve \eqnref{eq:relic} to find the coupling constant within the range $[10^{-9}, 1]$. LHS technique enables uniform and efficient sampling of this multidimensional input parameter space, avoiding clustering and undersampling in any region. To ensure comprehensive coverage of the parameter space, we apply logarithmic scaling to parameters spanning multiple orders of magnitude and linear sampling to those with narrower ranges. In total, we initially generate $30,000$ combinations of input parameters. For a subset of these points, no root was found for $\lambda_\phi$ within the range considered. After discarding these points, the final dataset comprises $21,577$ unique parameter combinations. Next, we structure our datasets with a four-dimensional input vector of $\{ \log m_\phi, \log \lambda_\phi, \log m_4, \omega_{c}\}$ and a corresponding output value $\log \theta$. The dataset is randomly shuffled and divided into independent training, validation, and test sets, comprising $16,520$, $2,360$, and $2,697$ samples, respectively. We normalize all three datasets using z-score standardization to facilitate efficient, optimal network training.

We employ a fully connected feed-forward NN. This network architecture comprises an input layer corresponding to the four input parameters $(\log m_\phi, \log \lambda_\phi, \log m_4, \omega_{c})$, multiple hidden layers with nonlinear activation functions, and a single output layer that produces a scalar output $\log \theta$. To capture the inherent nonlinearities in the parameter relationships, we employ the Rectified Linear Unit (ReLU) activation function in the hidden layers. The number of neurons per hidden layer is determined through hyperparameter optimization.

\begin{figure*}[!htbp]
\centering
	\includegraphics[width=1.0\textwidth]{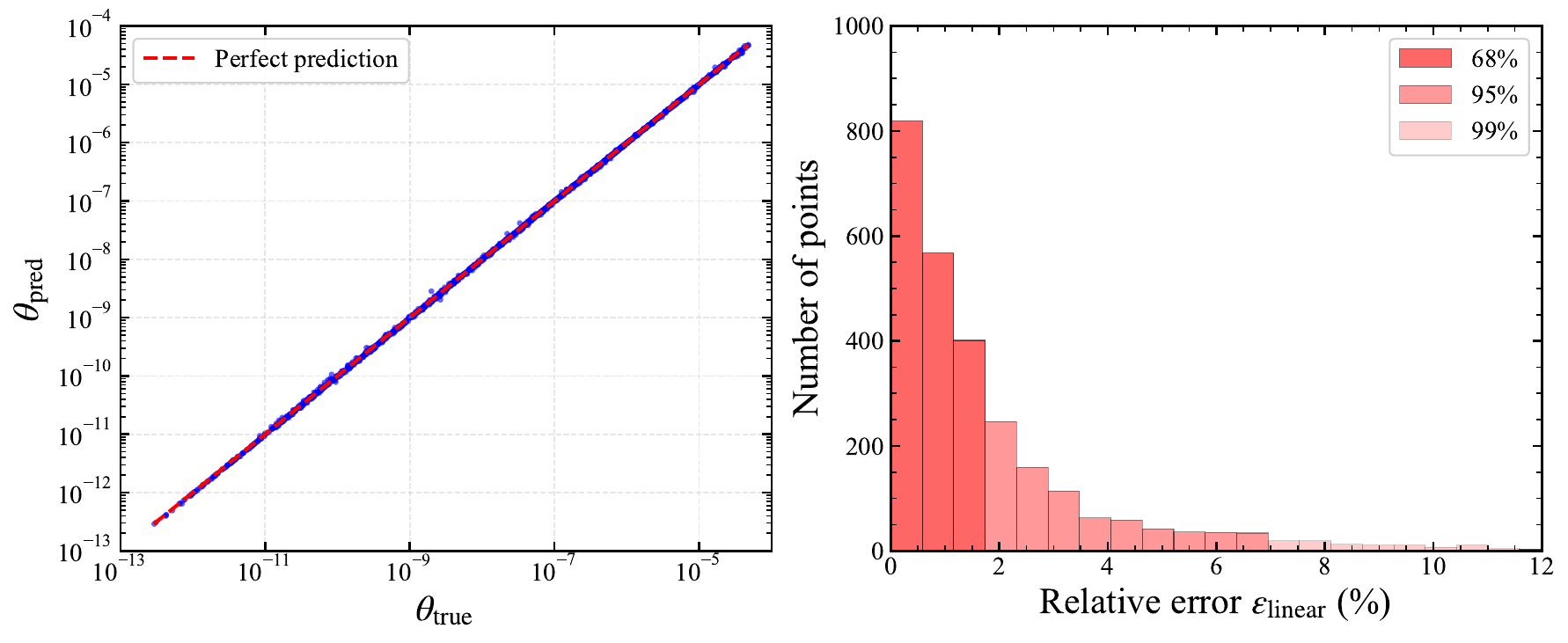}
\caption{Prediction performance of the NN-based emulator for the active-sterile neutrino mixing angle on the independent test dataset. {\bf Left panel:} Predicted values $\theta_{\rm pred}$ plotted against the actual values $\theta_{\rm true}$ as shown by the blue points. The red dashed line represent the actual $\theta$ values from the test data. {\bf Right panel:} Distribution of relative percentage errors $\varepsilon_{\rm linear}$ for the test data. The red shaded bands indicate regions with $68\%$, $95\%$, and $99\%$ of the total number of test data (from darkest to lightest, respectively).
}\label{fig:em_th_lin}
\end{figure*}

Next, we describe the training procedure and the optimization strategy. We use the Mean Squared Error (MSE) as a loss function for our regression model, defined as $\mathcal{L} = \frac{1}{N} \sum_{i=1}^{N} (\log_{10}(\theta_{\text{pred},i}) - \log_{10}(\theta_{\text{true},i}))^2,$ where $N$ denotes the number of samples. To minimize the loss function during the training process, we compare two optimization algorithms: Adam~\cite{kingma2014adam} and stochastic gradient descent (SGD)~\cite{robbins1951stochastic,ruder2016overview}. Training is performed with mini-batches of batch size $32$ and $64$, which balance computational efficiency with accuracy in gradient estimation. To avoid overfitting, we also implement an early-stopping mechanism with a maximum of $5000$ epochs. The early stop mechanism uses a step-decay learning rate, which enables faster convergence and helps avoid overfitting. The initial learning rate is set to $0.01$ and is reduced to $0.001$ if no improvement in validation loss is observed for $150$ consecutive epochs. A second reduction to $0.0001$ occurs after another $150$ epochs without improvement in validation loss. Training terminates if no further improvement in validation loss is observed for $200$ consecutive epochs. The final model is selected based on the weights corresponding to the minimum validation loss achieved during training. 

We perform a systematic hyperparameter search using the training procedure described above to find the optimal configuration of the NN model architecture. In this search, we fix the number of hidden layers at three and vary the number of neurons per hidden layer. We explore all combinations of $32$, $64$, and $128$ neurons per layer, resulting in 27 distinct network configurations. For each configuration, we repeat the training process with both mini-batch sizes ($32$ and $64$) and both optimization algorithms (Adam and SGD).
The best-performing model is selected based on the lowest validation loss achieved across all configurations. The best NN architecture is found to be $4 \rightarrow 32 \rightarrow 128 \rightarrow 32 \rightarrow 1\,$, which corresponds to an input layer with four parameters, three hidden layers with $32$, $128$, and $32$ neurons, respectively, and a single output.

The selected model is evaluated on the independent test data set. ~\figref{fig:em_th} (left panel) shows the predicted $\log_{10}(\theta_{\rm pred}$) plotted against the corresponding actual values $\log_{10}(\theta_{\rm true}$)), while \figref{fig:em_th_lin} (left panel) shows the same comparison in linear space ($\theta$). The emulator achieves high prediction accuracy across the entire parameter range. To quantify the prediction accuracy, we also compute the relative percentage error in both logarithmic and linear predictions as:
\begin{align}
\varepsilon_{\rm log} &= 100 \times \frac{|\log_{10}(\theta_{\rm true}) - \log_{10}(\theta_{\rm pred})|}{|\log_{10}(\theta_{\rm true})|}, \\
\varepsilon_{\rm linear} &= 100 \times \frac{|\theta_{\rm true} - \theta_{\rm pred}|}{\theta_{\rm true}}.
\end{align}
The distributions of the computed relative percentage errors are shown in the right panel of ~\figref{fig:em_th} and \figref{fig:em_th_lin}. The emulator predicts the mixing angle with less than $7\%$ error for $95\%$ times of the test samples, demonstrating excellent performance for the purposes of this work. 

In conclusion, the NN-based emulator developed in this work provides an accurate and computationally efficient surrogate model to predict active sterile mixing angles. The emulator is implemented using Python with the following libraries: PyTorch~\cite{paszke2019pytorch}, NumPy~\cite{harris2020array}, pandas~\cite{mckinney-proc-scipy-2010,reback2020pandas}, and Scikit-learn~\cite{scikit-learn}.

\end{document}